\documentclass{llncs}
\usepackage[english]{babel}  
\usepackage{hyperref,graphicx,alltt,xcolor,float}
\usepackage{color}
\usepackage{url}
\usepackage{listings}
\usepackage{multirow}
\begin{document}

\title{You Overtrust Your Printer}

\author{Giampaolo Bella\inst{1} \and Pietro Biondi\inst{1,2}}

\institute{Dipartimento di Matematica e Informatica, Universit\`a di Catania, Italy
\email{giamp@dmi.unict.it}
\and
Istituto di Informatica e Telematica - Consiglio Nazionale delle Ricerche
\email{pietro.biondi94@gmail.com}}

\maketitle

\begin{abstract}
 Printers are common devices whose networked use is vastly unsecured, perhaps due to an enrooted assumption that their services are somewhat negligible and, as such, unworthy of protection. This article develops structured arguments and conducts technical experiments in support of a qualitative risk assessment exercise that ultimately undermines that assumption. Three attacks that can be interpreted as post-exploitation activity are found and discussed, forming what we term the Printjack family of attacks to printers. Some printers may suffer vulnerabilities that would transform them into exploitable zombies. Moreover, a large number of printers, at least on an EU basis, are found to honour unauthenticated printing requests, thus raising the risk level of an attack that sees the crooks exhaust the printing facilities of an institution. There is also a remarkable risk of data breach following an attack consisting in the malicious interception of data while in transit towards printers. Therefore, the newborn IoT era demands printers to be as secure as other devices such as laptops should be, also to facilitate compliance with the General Data Protection Regulation (EU Regulation 2016/679) and reduce the odds of its administrative fines.
\end{abstract}

\section{Introduction}
The era of the Internet of Things (IoT) has only just begun \cite{introIoT}. Electronic devices of various nature are starting to be endowed with WiFi modules that connect them to the local network. This revolution concerns 
both private contexts, such as peoples' houses and devices, as well as professional contexts, such as peoples' (institutional) workplaces. For example, doors, gates, power switches, heating systems, water timers, blood pressure monitors and many other devices can be connected and operated via a remotely connected computer. 

However, the comfort of using such smart equipment, for example, via a smartphone while the user sits on her sofa, comes at the cost of a drastically increased risk of remote, malicious activity by some attacker. A remarkable example published these days shows an after-market, Android 6.0 car radio suffering a simple vulnerability: an unauthenticated, root-level, remote access \cite{candycream}. In consequence, Costantino and Matteucci tailor a post-exploitation script that packages CAN bus traffic to vandalize the odometer of the car. Their attack assumes the attacker to have gained access to the in-vehicle network, for example through the diagnostic OBD2 port or by exploiting a vulnerability that the e-call box system of the car may have. While cars become more and more interconnected with the IoT, the researchers' assumption gets more and more realistic by the minute.

Our work concentrates on printers, devices that are still tremendously used in every context, despite a perceivable quest for a paperless revolution. We find out that a large number of printers is publicly exposed over the Internet and that, at the same time, data sent to printers is often unsecured in the sense that a printer may honour unauthenticated print jobs --- remarkably, even if these are sent from a remote network thanks to the printer being visible over the Internet. Moreover, such jobs do not transfer user data confidentially, namely data will traverse the local network in the clear towards the printer. 

As a consequence of the lack of authentication, printers may suffer vulnerabilities that may turn out to be exploitable even remotely; moreover, those printers may be put at stake by (local or remote) jobs that are sent repeatedly with a malicious aim. As a consequence of the lack of confidentiality, should an attacker get on any node of the local network (by exploiting a vulnerability of that node), he could intercept the print jobs sent by a legitimate user, understand and abuse them causing a data breach. If this happens in the institution where the user is employed, and the intercepted print jobs carry anyone's personal data, then EU Regulation 2016/679, the General Data Protection Regulation (GDPR) \cite{gdpr} states that the institution may be severely fined, as we shall see below.

Our attack resembles the mentioned one to the odometer of the car because the CAN bus also lacks authentication and confidentiality measures. In the car, the attacker leverages CAN bus traffic being in the clear to understand how the odometer would react to specific CAN frames; he can then bombard the odometer with chosen frames without any authentication hurdle. By contrast, our attack to printers appears to be more multi-faceted than that, and in fact shapes up as three different attacks, the Printjack (which stands for \emph{print}er hi\emph{jack}ing) family of attacks to printers.

\section{Summary of the contributions}\label{sec:contribution}
This article evaluates some of the possible consequences of the use of raw 9100 port printing. As a start, we used a free student account on Shodan, the search engine for the IoT \cite{shodan}, to determine how common the bad practice of exposing public IP addresses over the Internet with a responding 9100 port is. We were surprised to find out almost three thousand occurrences in the authors' country, which we obtained by querying Shodan with:
\begin{alltt}
   port:9100 country:"IT"
\end{alltt}
By varying the country identifier, we continued to obtain unexpected results. Table \ref{tab:pil} sorts European countries by their 2018 Gross Domestic Profit (GDP) \cite{gdp} and reports the number of IPs with open 9100 port that are exposed over the Internet from that country, according to the data we gathered through our Shodan queries.
For example, it turns out that the country with highest GDP, Germany, also exposes the highest number of devices.

\begin{table}[h]
\begin{center}
	\begin{tabular}{|c|c|c|}
		\hline 
		\textsc{GDP} & \textsc{Country} & \textsc{IPs with responding 9100 port} \\ 
		\hline 
		1 & Germany & 12.891 \\ 
		\hline 
		2 & Russia & 9.737 \\ 
		\hline 
		3 & United Kingdom & 6.349 \\ 
		\hline 
		4 & France & 6.634 \\ 
		\hline 
		5 & Italy & 2.787 \\ 
		\hline 
		6 & Spain & 2.088 \\ 
		\hline 
		7 & Turkey & 835 \\ 
		\hline 
		8 & Poland & 1.425 \\ 
		\hline 
		9 & Netherlands & 4.934 \\ 
		\hline 
		10 & Switzerland & 624 \\ 
		\hline 
	\end{tabular}
\end{center}
\caption{IPs with responding 9100 port per country, sorted by country's GDP}\label{tab:pil}
\end{table}

We interpret the high numbers noted above as a widespread, publicly available, potential vulnerability. Of course, we refrained from attempting to connect to those devices for ethical reasons. It must be noted, however, that, although one can configure any service behind any port, raw printing is the default service for 9100 port, hence it is likely to be left as is. These findings give strength to the remaining contributions of this article.
We define the Printjack family of attacks to printers as post-exploitation activity following the reported vulnerability:
\begin{itemize} 
	\item Printjack 1 attack: zombies for traditional DDoS (\S\ref{sec:zombie})
	\item Printjack 2 attack: paper DoS (\S\ref{sec:paper})
	\item Printjack 3 attack: privacy infringement (\S\ref{sec:privacy})
\end{itemize}

We evaluate each attack using a qualitative risk assessment approach based upon the ISO/IEC 27005:2018 standard \cite{iso27005}. In particular, we  develop structured arguments and conduct technical experiments to evaluate the \emph{likelihood} and the \emph{impact} of each attack with the aim of calculating the \emph{risk level} of the attack. The calculation is based on Table \ref{tab:ra}.

\begin{table}[h]
  \begin{center}
    \begin{tabular}{|c|r||c|c|c|c|c|}
      \hline
      \multicolumn{7}{|c|}{\textbf{risk impact}} \\ \hline \hline
      \multirow{6}{*}{\textbf{\rotatebox[origin=l]{90}{risk likelihood}}} & \multicolumn{0}{c||}{} & \textsc{minor} & \textsc{moderate} & \textsc{major} & \textsc{severe} & \textsc{catastrophic} \\ \cline{2-7}
      & \textsc{rare} & \textsc{low} & \textsc{low} & \textsc{low} & \textsc{low} & \textsc{low} \\ \cline{2-7} 
      & \textsc{unlikely} & \textsc{low} & \textsc{low} & \textsc{medium} & \textsc{medium} & \textsc{medium} \\ \cline{2-7} 
      & \textsc{possible} & \textsc{low} & \textsc{medium} & \textsc{medium} & \textsc{high} & \textsc{high} \\ \cline{2-7} 
      & \textsc{likely} & \textsc{low} & \textsc{medium} & \textsc{high} & \textsc{high} & \textsc{extreme} \\ \cline{2-7} 
      & \textsc{almost certain} & \textsc{low} & \textsc{medium} & \textsc{high} & \textsc{extreme} & \textsc{extreme} \\ \hline
    \end{tabular}
  \end{center}
\caption{Evaluation of the risk level according to ISO/IEC 27005:2018}\label{tab:ra}
\end{table}

To our own surprise, all Printjack attacks are found to bear risk level {\sc HIGH}. Despite the inherent subjectivity of the risk assessment exercise, we are confident that it synthesises our arguments and experiments correctly as well as profitably.

This manuscript continues with a discussion of the related work (\S\ref{sec:related}) and concludes by deriving lessons learned and outlining possible fixes (\S\ref{sec:concl}).

\section{Printjack attack 1: zombies for traditional DDoS}\label{sec:zombie}
It is well known that Denial of service (DoS) perhaps is the most severe attack in the modern Internet era. The implicit loss caused by an unresponsive service can be enormous, and figures get continuously updated \cite{ddos}. The distributed version of this attack (DDoS) sees an attacker operate a \emph{Command and Control} server to administer a number of infected computers that are normally called \emph{zombies} or \emph{botnets}. 

One of the implications of the IoT era is that zombies could be farmed from any interconnected device with some computational power, provided it suffers some vulnerability that would enable its remote hijacking. A recent scandal saw more than a million cameras zombied to mount a massive DDoS \cite{camerasvice}. It is clear that the inherent performance of each zombie, which may be relatively low, is offset by the huge number of available devices. 

Turning the focus back to printers, it can be noted that there exist a number of documented vulnerabilities on various printers, which can be found
on the Common Vulnerabilities and Exposures (CVE) database by the MITRE \cite{homemitre}. These observations motivate a daunting research question: how significant is the risk that worldwide printers get exploited to mount a massive DDoS attack? We argue that risk to be high hence worthy of mitigation, and provide supporting evidence for this argument below.

\subsection{Supporting evidence}
We take a stab at answering the question posed above by addressing the risk that a DDoS attack sourced from printers would take place. This can be done, in turn, by means of a qualitative risk assessment approach. There are a number of CVEs about printer vulnerabilities, precisely 179 can be found by querying the CVE database with keyword ``printer'' \cite{cveprinter} and 77 by querying it with keyword ``printers'' \cite{cveprinters}, totalling 223 by adding up and removing intersections. In particular, we observe that a few dozens of these allow for the remote execution of arbitrary commands or code. For example, CVE-2014-3741 ``\emph{allows remote attackers to execute arbitrary commands via unspecified characters in the lpr command}'' \cite{unavulnprinter}.

We contend that these findings, in combination with the potential for zero-day attacks, raise the attack likelihood to {\sc possible}. 
Similarly, the widespread reachability of the 9100 port on real printers we noted above, of nearly 50K units only across the top ten wealthiest EU countries  (Table~\ref{tab:pil}), justifies a {\sc catastrophic} attack impact.  
According to Table~\ref{tab:ra}, the assessed likelihood and impact of the Printjack 1 attack lead to a {\sc high} risk level. A risk of this level must be mitigated as soon as possible.

\section{Printjack attack 2: paper DoS}\label{sec:paper}
M\"uller et al. exhibit a proof of concept on how to mount a DoS on printers \cite{expnetprinters}. It keeps the PostScript interpreter of the printer busy forever by means of an infinite loop (based on an empty instruction and an empty exit condition). The researchers confirmed this attack on all their twenty tested printers but the HP LaserJet M2727nf, which automatically rebooted after ten minutes.

We note that raw port 9100 printing can be exploited to potentially exhaust the printing facilities of an institution. It can be done by abusing via the  9100 port any printer that becomes known through its IP address. An attacker would send repeated print jobs till the victim printer runs out of paper from all its paper trays. Looping on all institutional printers would then complete the attack. We conjecture that, in practice, a legitimate print attempt in front of a printer that processed all available paper (by printing something on each sheet and making it useless) would lead the employees to reload some paper trays. As an extreme, the institution would run out of paper
should the reloads persist before the attack is found and removed.

The Printjack 2 attack is of socio-technical nature because it is rooted in people's most obvious reaction to an aborted print attempt of theirs. It would be worth conducting field studies to verify our conjecture that people would feed their printers more and more paper unless they get their printout. This is beyond our present aims; by contrast, we provide a proof of concept implementation of the technical part below.

\subsection{Supporting evidence}
The technical part of the Printjack 2 attack can be easily implemented in Python as shown in Table \ref{tab:script}. By looking at it from the inside out, we see a loop that sends each line, stored in {\tt textlines}, of a bot ASCII file {\tt bot.txt}, stored in {\tt textfile}, to a printer for a thousand times. The bot file could contain anything that the attacker may want to print in order to process and spoil paper sheets. The printer is identified via its IP address, and a socket connects to its 9100 port. The outermost loop ranges on the target IP addresses, which are read from file {\tt IPs.txt}. 

\begin{table}[h]
\begin{lstlisting}
 import socket
 f = open("IPs.txt", "r") #file containing IPs of target printers
 lines = f.readlines()
 for ip in lines:
  	textfile = open("bot.txt", "r") #ascii file to be printed
  	textlines = textfile.readlines()
  	for count in range(0,1000): #number of print jobs
    		s = socket.socket()
    		s.connect((ip, 9100))
    		for line in textlines:
      			s.send(line+"\n")
    		s.close()
\end{lstlisting}
\caption{Python script for our paper DoS attack}\label{tab:script}
\end{table}

\begin{figure}[h]
\begin{center}
\includegraphics[scale=0.42]{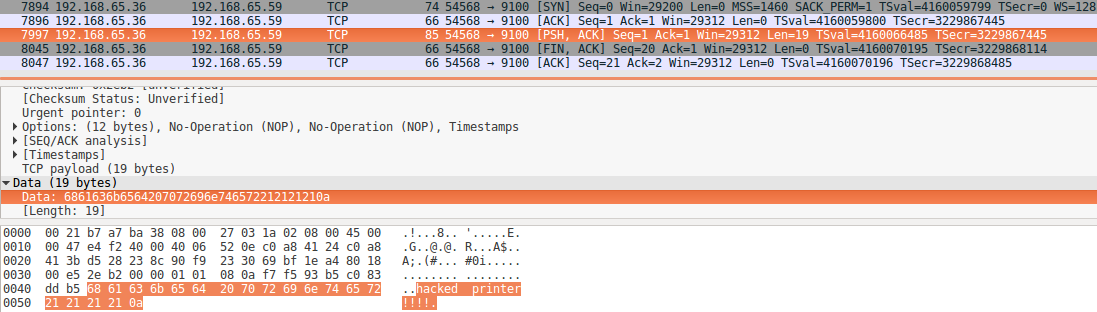}
\end{center}
\caption{The Printjack 2 attack monitored via Wireshark}\label{fig:sniffing1}
\end{figure}

We run our script on our institutional LAN. More in detail, we launched it from within the network, precisely from private IP address 192.168.65.36, towards a target printer of IP address 192.168.65.59. The printer exhausted its available paper by marking each sheet with the test phrase ``hacked printer!!!!''. Feeding it more paper would of course continue the paper abuse because the stated one thousand threshold had not been reached yet. We had to reset the printer manually to terminate the ignominy. Our experiment can be confirmed by observing the network traffic as sniffed by Wireshark \cite{wireshark}. The screenshot in Figure \ref{fig:sniffing1} highlights the appropriate TCP connection and the test phrase.

Reproducing the Printjack 2 attack on a large scale, by targeting remote printers, does not seem difficult although we have obviously not tried that for ethical reasons. In a practical scenario, file {\tt IPs.txt} could be built by appropriately querying Shodan. 
We decided to query the EU country with the highest GDP, that is, Germany. Therefore, our query was:
\begin{alltt}
   port:9100 country:"DE"
\end{alltt}

The results can be conveniently exported as a CSV file by paying some Shodan credits. We decided to pay one Shodan ``export credit'', which obtained us ten thousand entries. Our student account granted us one hundred Shodan credits for free, so it is remarkable that it was free 
to obtain that much information and that it would still be free to obtain (much) more. For the sake of demonstration, Table \ref{tab:shodanip} shows public information, a small excerpt of the 2.3MB file that Shodan built for us to download.

\begin{table}[h]
\begin{center}
	\begin{tabular}{|c|c|}
		\hline 
		\textsc{IP} & \textsc{PORT} \\ 
		\hline 
		87.156.104.144 & 9100 \\ 
		\hline 
		79.231.20.111 & 9100 \\ 
		\hline 
		141.24.208.236 & 9100 \\ 
		\hline 
	\end{tabular}
\end{center}
\caption{An excerpt of the 10000 entry file with target IPs exported from Shodan}\label{tab:shodanip}
\end{table}

In conclusion, the public availability of remote 9100 printer ports noted above (\S\ref{sec:contribution}), which can be practically leveraged by building tables such as Table \ref{tab:shodanip}, supports the claim that this attack is reproducible remotely on a large scale.

It is worth to qualitatively risk-assess also the Printjack 2 attack. Because our conjecture on the socio-technical part is yet unverified, we contend a {\sc possible} attack likelihood.
However, it is evident that the attack impact is {\sc severe}, hence the resulting risk level is {\sc high}.

\section{Printjack attack 3: privacy infringement}\label{sec:privacy}
The treatment unfolded thus far emphasises that print jobs may be sent in the clear.
Suppose that an attacker Mallory sits on the same network as some target employee Alice. This scenario is normally addressed as an \emph{insider threat}. We note that whenever Alice sends a print job in the clear, Mallory could carry out a Man In The Middle (MITM) attack and eavesdrop the printed material, a clear infringement of Alice's privacy. Mallory could misbehave further, by publishing the intercepted material anonymously on the Internet, and produce a data breach. 

A similar attack scenario sees a remote attacker Eve exploit one vulnerability into Alice's institutional network. It is state of the art to protect critical resources such as servers and databases by means of (strong) authentication. So, because Eve operates on the one node affected by the assumed vulnerability, those critical resources remain protected. By contrast, Eve could still perform the print job eavesdropping described above. Because printing is still common practice today, we cannot fully justify why data stored on a server would normally be protected and, by contrast, data sent off for printing would not.

The impact of such events would be very serious in our epoch, at least in the EU, where citizens' data protection is regulated by the GDPR. With its 99 articles, the regulation empowers people with a number of rights to be exercised over their personal data as hosted by any data controller institution. The GDPR also stresses the responsibilities of the controller, for example article 5 paragraph 2 states that ``\emph{The controller shall be responsible for, and be able to demonstrate compliance with, paragraph 1 (`accountability').}'', with the mentioned paragraph 1 setting the requirement, among others, that data be ``\emph{processed in a manner that ensures appropriate security of the personal data, including protection against
unauthorised or unlawful processing and against accidental loss, destruction or damage, using appropriate technical
or organisational measures (`integrity and confidentiality').}''. Moreover, article 83 threatens ``\emph{administrative fines
up to 20 000 000 EUR, or in the case of an undertaking, up to 4 preceding financial year, whichever is higher.}''.

Alice's institution has a great lot to worry about, equally because of Mallory's misconduct and because of Eve's.

\subsection{Supporting evidence}
Evidence seen in Figure \ref{fig:sniffing1} is valid also in the threat models embodied respectively by Mallory and Eve. In such cases, the visible traffic could be interpreted as Alice's, clearly intelligible, private data that Alice sent off for printing in a file intercepted by the attacker.

\begin{figure}[h]
\begin{center}
\includegraphics[scale=0.45]{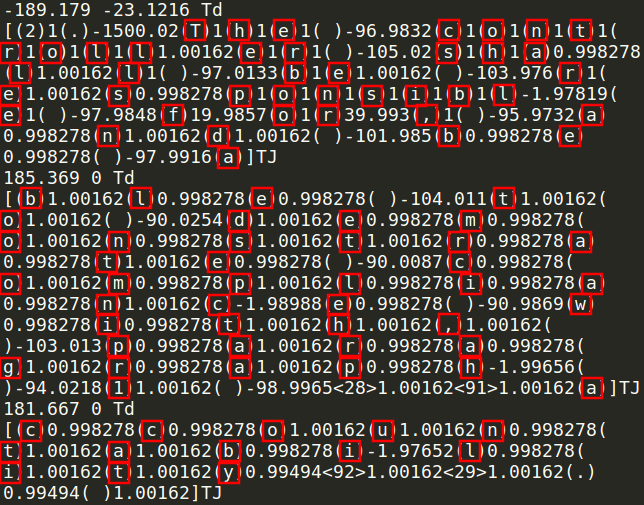}
\end{center}
\caption{Sniffing a PDF file (containing the GDPR) as printed from Linux}\label{fig:privacy2}
\end{figure}

To inform a qualitative risk assessment upon the Printjack 3 attack, conducted in the following, we remark that raw port 9100 printing is massively used worldwide. For example, we observe that it is the default print method that the Common UNIX Printing System (CUPS) leverages, and that CUPS is vastly used in modern   Linux distributions and Apple systems. As a demonstration, we used Ettercap \cite{ettercap} to interpose through sender and printer, then Wireshark to intercept the PDF file of the GDPR as from its official URL \cite{gdpr}. The outcome is intelligible with some decoding. The excerpt in Figure \ref
{fig:privacy2} highlights in red the mentioned text of article 5 as intercepted over a print job sent from a Fedora 28 machine. It would be easy to implement a pretty-priting script.

Our print job sniffing experiments took a different course when the jobs were sent from an updated Windows 10 machine. While M\"uller et al. claim that Microsoft Windows printing architecture uses raw port 9100 printing by default \cite{expnetprinters}, our sniffing experiments yielded no comprehensible material. Although more experiments are needed to fully scrutinise this scenario, it would seem that 9100 no longer is the default printing port on Windows, thus supporting the claim that printing is more secure from Windows machines at present than from other systems.

\begin{figure}[h]
\begin{center}
\includegraphics[scale=0.45]{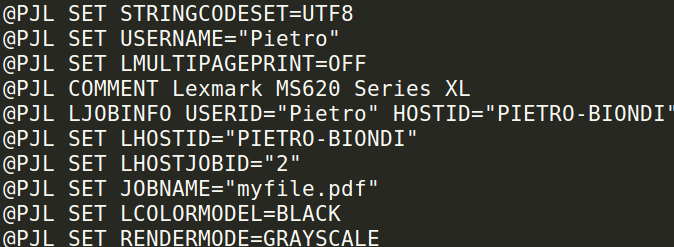}
\end{center}
\caption{Sniffing the metadata of a PDF file on Windows}\label{fig:privacy1}
\end{figure}

Nevertheless, we succeeded in intercepting the print job metadata on Windows. Figure
\ref{fig:privacy1} shows the metadata intercepted over port 65002, precisely fields {\tt USERNAME}, {\tt USERID}, {\tt HOSTID}, {\tt JOBNAME} as well as the printer model. Although this is less intrusive than accessing the contents of the printed file, it still counts as a data breach at least for the meaningful association of the file name to the user name. This claim rests on the socio-technical assumption that people give files meaningful names.

In light of the above experiments and collected evidence, we argue the likelihood of the Printjack 3 attack to be {\sc likely} and its impact to be {\sc severe}, hence its resulting risk level is {\sc high}.

\section{Related work}\label{sec:related}
The most eminent piece of research in the areas of printer security and privacy is due to M\"uller et al. \cite{expnetprinters}. They conduct a full-breadth vulnerability assessment and penetration testing session over a range of twenty commercial printers, comparing and contrasting a number of attacks on each of them. Their work is the first to note that raw 9100 port printing may be risky.

It must be mentioned that the work by M\"uller et al. also led to the development of the Printer Exploitation Toolkit (PRET), which is available on GitHub \cite{pret}. However, we report that the technical parts of the Printjack family of attacks discussed above did not work using the tool against our main testbed printer, a Lexmark MS620.
PRET is the newest and best developed of a small bunch of tools \cite{toolwiki}, which could not be used successfully for our purposes either.

In the same year when the research findings by M\"uller et al. appeared, 2017, they were sided with breaking news reporting large-scale printer hacking somewhat for fun \cite{printervice}, and the news was reiterated in 2018. The technical foundations behind the news remain vague, of course. Moreover, it is not obvious to what extent the research findings inspired the events outlined in the news and, vice versa, whether the news partly ignited the researchers' investigations.

\section{Conclusions}\label{sec:concl}

There is awareness that the IoT era has only just began, and more and more devices will be connected to the Internet over time.
The Printjack family of attacks demonstrates that printers are routinely \emph{not} configured and used with security and privacy in mind. Although the IoT revolution has driven the security-and-privacy eye that we have casted at printers, it must be noted that printers started to be networked even before the inception of the IoT era, and this makes our findings all the more surprising.

In conclusion, we remark that the {\sc high} risk level of the Printjack 1 attack was mostly determined by its impact rather than by its likelihood. The Printjack 2 attack could be carried out both from a local attacking machine or from a remote one if the target printers are exposed over the Internet. By contrast, the Printjack 3 attack can only be mounted against a user and a printer only if the attacking machine is local to them, hence the attacker must have exploited some vulnerability over a (node of) the network.

Well beyond the technicalities of the attacks lies a clear lesson learned. Printers ought to be secured equally as other network devices such as laptops normally are. A few appropriate security measures can be envisaged. For example, if user access to a laptop is normally authenticated, then so should be user access to the web-server-based admin panel of a printer, which often allows, for example, printer reset, printer name change, access to list of printed file names, etc. Similarly, remote connection to a port of a laptop will be bound to authentication to some daemon and, likewise, sending a print job should require an extra level of authentication to the printer. 

Analogous considerations apply to data normally being encrypted while in transit between computers; this leads to the idea of encrypting print jobs too. All these specifications could be implemented, for example, by enabling IPSec-only connections to printers, a feature that inexpensive printers currently offer. The reason why this feature does not seem commonly used may boil down to the traditional usability imbalance at the expenses of protection. Since appropriate technology is available to mitigate the risks of the Printjack family of attacks to printers, the biggest effort ahead of us seems to be the training of users to bear security and privacy measures also through their routine printing tasks.

\section*{Acknowledgements}
We are indebted to Gianpiero Costantino and Ilaria Matteucci for arousing innumerable inspiring discussions.

\bibliographystyle{splncs}
\bibliography{iotbiblio}

\begin{thebibliography}{10}

\bibitem{introIoT}
Shemshadi, A., Sheng, Q.Z., Qin, Y., Sun, A., Zhang, W.E., Yao, L.:
\newblock Searching for the internet of things: where it is and what it looks
  like.
\newblock Personal and Ubiquitous Computing \textbf{21} (2017)  1097--1112

\bibitem{candycream}
Costantino, G., Matteucci, I.:
\newblock {CANDY CREAM} - ha{C}king infot{A}i{N}ment an{D}roid s{Y}stems to
  {C}ommand inst{R}ument clust{E}r via c{A}n data fra{M}e.
\newblock In: Proceedings of the 17th IEEE International Conference on Embedded
  and Ubiquitous Computing. IEEE EUC '19 (2019) In press.

\bibitem{gdpr}
Union, E.:
\newblock {{G}eneral {D}ata {P}rotection {R}egulation({EU} {R}egulation
  2016/679)}.
\newblock
  \url{https://eur-lex.europa.eu/legal-content/EN/TXT/PDF/?uri=OJ:L:2016:119:FULL}
  (2016)

\bibitem{shodan}
search engine for the Internet~of Things, T.:
\newblock {Shodan}.
\newblock \url{https://www.shodan.io/} (2019)

\bibitem{gdp}
Wikipedia:
\newblock {European states by GDP}.
\newblock
  (\url{https://en.wikipedia.org/wiki/List_of_sovereign_states_in_Europe_by_GDP_(nominal)})

\bibitem{iso27005}
{International Organization for Standardization}:
\newblock {Information technology -- Security techniques -- Information
  security risk management}.
\newblock \url{https://www.iso.org/standard/75281.html} (2018)

\bibitem{ddos}
Sirbu, M.:
\newblock Security concerns in a 5g era: are networks ready for massive ddos
  attacks?
\newblock
  \url{https://www.scmagazineuk.com/security-concerns-5g-era-networks-ready-massive-ddos-attacks/article/1584554}
  (2019)

\bibitem{camerasvice}
Vice:
\newblock {How 1.5 Million Connected Cameras Were Hijacked to Make an
  Unprecedented Botnet}.
\newblock
  \url{https://www.vice.com/en_us/article/8q8dab/15-million-connected-cameras-ddos-botnet-brian-krebs}
  (2016)

\bibitem{homemitre}
MITRE:
\newblock {Home page}.
\newblock \url{https://cve.mitre.org/} (2019)

\bibitem{cveprinter}
MITRE:
\newblock {CVE}-printer.
\newblock \url{https://cve.mitre.org/cgi-bin/cvekey.cgi?keyword=printer} (2019)

\bibitem{cveprinters}
MITRE:
\newblock {CVE}-printer.
\newblock \url{https://cve.mitre.org/cgi-bin/cvekey.cgi?keyword=printers}
  (2019)

\bibitem{unavulnprinter}
NIST:
\newblock {CVE}-2014-3741.
\newblock \url{https://nvd.nist.gov/vuln/detail/CVE-2014-3741} (2014)

\bibitem{expnetprinters}
{Müller}, J., {Mladenov}, V., {Somorovsky}, J., {Schwenk}, J.:
\newblock Sok: Exploiting network printers.
\newblock In: 2017 IEEE Symposium on Security and Privacy (SP). (2017)
  213--230

\bibitem{wireshark}
Wireshark:
\newblock {Wireshark project}.
\newblock \url{https://www.wireshark.org/} (2019)

\bibitem{ettercap}
Ettercap:
\newblock {Ettercap project}.
\newblock \url{https://www.ettercap-project.org/} (2019)

\bibitem{pret}
GitHub:
\newblock {Printer Exploitation Toolkit}.
\newblock \url{https://github.com/RUB-NDS/PRET} (2018)

\bibitem{toolwiki}
Muller, J.:
\newblock {Printer Tool Wiki}.
\newblock \url{http://hacking-printers.net/wiki/index.php/Main_Page} (2017)

\bibitem{printervice}
Vice:
\newblock {This Teen Hacked 150,000 Printers to Show How the Internet of Things
  Is Shit}.
\newblock
  \url{https://www.vice.com/en_us/article/nzqayz/this-teen-hacked-150000-printers-to-show-how-the-internet-of-things-is-shit}
  (2017)

\end{thebibliography}

\end{document}